%% file: Driver.tex
\documentclass[11pt,psfig,epsfig]{article}
\usepackage{amssymb,latexsym}
\usepackage{amsmath}
\usepackage{graphicx}
\usepackage{url}

\begin{document}
\bibliographystyle{ieeetr}
\input{Header}
\input{Intro}

\input{KnownModel}

\input{PR_Model}
\input{Conclus}

\bibliography{Rumor}
\end{document}

%% file: Header.tex
\title{A deterministic mathematical model for the spread of two rumors}
\author{ }
\author{Ren\'e Escalante\thanks{Departamento de C\'omputo Cient\'{\i}fico y Estad\'{\i}stica,
 Divisi\'on de Ciencias F\'{\i}sicas y Matem\'aticas, Universidad Sim\'on
Bol\'{\i}var, Ap. 89000, Caracas, 1080-A, Venezuela
(rescalante@usb.ve, markodehnal@gmail.com).} \thanks{This author was partially supported by the Decanato de
Investigaci\'on y Desarrollo (DID) at USB.} \thanks{Corresponding author.} \and Marco Odehnal\footnotemark[1]
}
 \date{September 6, 2017}
\maketitle
\begin{abstract}
In this paper we propose a deterministic mathematical model that attempts to explain the 
propagation of a rumor using SIRS type epidemiological models with temporary immunity and nonlinear incidence rate.
 In particular, we speculate about the dissemination of information when the so-called ``complex networks" are used.
The effect of introducing a second rumor, inspired by a vaccination model, in the same population of individuals, which will try to counteract the effect of the original rumor, is 
 studied. That is a situation that occurs frequently in communities, when a rumor is counteracted by a contrary
 information or news, which behaves in the same way as a rumor.
 Furthermore, qualitative analysis and numerical experimentation of the dynamic model are performed.
We corroborate that the dynamics of spreading rumors show similar behavior to that found in the dynamics of an infectious disease. \\ [2mm] {\bf Keywords:} Mathematical model; Spreading rumor; SIRS model; Vaccination model; Epidemic. 
\end{abstract}
\newpage

%% file: Intro.tex
\section{Introduction and Preliminaries} \label{Introd}
 Information spreads in ways that resemble the transmission dynamics of viruses. In order to help emphasize the similarity to epidemic models, the customary variable names in epidemiology will be employed throughout. The medium through which the rumors are transmitted is also an important factor in the dynamics of spreading. Actually, computers are constantly used and are great means for transmitting information as well as rumors. This led to consider a particular form of epidemic models, which here we prefer to call the propagation rumor (PR) model.

At first glance it seems that the PR model is an example of logistical behavior, because, initially, only a small fraction of the population may know some rumor. These individuals pass the rumor on to their neighbors. The cycle repeats and soon there are so many people who have heard the rumor that it becomes difficult to find someone who has not heard it. The number of people hearing the rumor for the first time begins to level off creating an S-shaped curve \cite{Pearce}. However, the phenomenon of the spread of rumors is somewhat more complex.

The standard PR model was introduced in 1965 by Daley and Kendall \cite{DaleyKendall:65} (DK model). Also, the modeling of rumor propagation has been proposed by other authors such as Rapoport (1948) and Bartholomew (1967) (see \cite{DaleyGani:01} and references therein), and Zanette (2001) \cite{Zanette:01}. These approaches are based on stochastic processes. 
 In this paper we propose a deterministic approach, which draws from the SIRS epidemiological models along with the introduction
 of a second rumor
 (of different nature than the first) in the susceptible population. 

  We formulate our proposal as a compartmental model \cite{BauerCast:12}, with the population being divided
  into compartments and with assumptions
  about the nature and time rate of transfer from one compartment to another.
  We describe a model for propagation of rumors, acting on a sufficiently rapid time scale such that demographic effects
  of a population may be ignored.
  Our model is concerned with the spread of a rumor through individuals of a given population, with a total population size
  of $n$ individuals.


An instructive approach is to treat the problem by qualitative methods. We start with the classical Kermack-McKendrick epidemic model (KM model) \cite{KermMcKen:27}

\begin{equation} \label{KMmodel}
\left\{ \begin{array}{lll}
\frac{ds(t)}{dt} & =  & -\beta s(t)i(t), \\[1.5mm]
\frac{di(t)}{dt} & = & \beta s(t)i(t) - \alpha i(t), \\[1.5mm]
\frac{dr(t)}{dt} & = & \alpha i(t),
 \end{array}
 \right.
\end{equation}
where $s(t)$, $i(t)$, and $r(t)$ denote, respectively, the numbers of susceptible, infectious, and removed individuals at time $t$,
and the parameters $\beta >0$ and $\alpha >0$ are known as the infection and removal rates.
This model, also called SIR model, forms the basis of all epidemiological models, which virtually omits the population of parasites (population of rumors) from direct consideration and assumes that the sizes of the compartments are large enough that the mixing of members is homogeneous. It is based on the following assumptions: an average member of the population makes sufficient contact to transmit the rumor to $\beta n$ others per unit time ({\it mass action law}), those infected by a rumor leave the infective class
at $\alpha i$ rate per unit time, and the time scale of the rumor is much faster than the time scale of births and deaths, so that demographic effects on the population may be ignored and the population remains constant (i.e., equal to $n=s+i+r$). We need to add to the KM model supplementary conditions: $s(t_0)=s_0>0$, $i(t_0)=i_0>0$, and $r(t_0)=0$.

Next, we shall carry out this procedure on a slightly more general case, allowing for a loss of immunity that causes recovered individuals to become susceptible again (i.e., {\it temporary immunity}). In our case, individuals which have previously heard the rumor, and were no longer interested in spreading it to other individuals (i.e., population of type $r$), suddenly, for some reason, become willing to spread rumors immediately. It will be assumed that this takes place at a rate proportional to the population of type $r$. Thus the equations become
\begin{equation*}
\left\{ \begin{array}{lll}
\frac{ds(t)}{dt} & =  & -\beta s(t)i(t) + \gamma r(t), \\[1.5mm]
\frac{di(t)}{dt} & = & \beta s(t)i(t) - \alpha i(t), \\[1.5mm]
\frac{dr(t)}{dt} & = & \alpha i(t) - \gamma r(t),
 \end{array}
 \right.
\end{equation*}
with a proportional rate $\gamma $ of loss of immunity.
 This model is called an SIRS model since removed individuals can return to class $s$ (i.e., a rate of transfer from $r$
 to $s$ is added to an SIR model \cite{BauerCast:12}).
 Since $n'=(s+i+r)'=0$, the total population size $n$ is constant and $r=n-s-i$. So, we may express the previous model by
 the system
 \begin{equation} \label{SIRSmodel}
\left\{ \begin{array}{lll}
\frac{ds(t)}{dt} & =  & -\beta s(t)i(t) + \gamma (n-i(t)-s(t)), \\[1.5mm]
\frac{di(t)}{dt} & = & \beta s(t)i(t) - \alpha i(t).
 \end{array}
 \right.
\end{equation}
  However, more complicated compartmental structures are possible; for example, there are SEIR and SEIS models, with an exposed period between being infected (i.e., to hear the rumor) and becoming infective (i.e., spreading the rumor) \cite{BauerCast:12}.
  Clearly, from system~(\ref{SIRSmodel}) there is a rumor-free equilibrium corresponding to $i=0$, $s=n$. In this equilibrium the whole population is healthy (but susceptible) and the rumor eventually disappears. The other stable state is $s=\alpha /\beta$,
  $i=\gamma (n - s)/(\alpha + \gamma)$. Here, for $i$ to be positive, necessarily, $n > s$. Since $s=\alpha /\beta$,
  the rumor remains in the population provided the total population $n$ exceeds $\alpha /\beta$. This is, whenever
  ${\cal R}_0 = n\beta / \alpha > 1$, where ${\cal R}_0$ has been called the {\it basic reproduction number} (see, for example, \cite{DriesWat:02}). This important threshold effect was discovered by Kermack and McKendrick (i.e.,
  population must be ``large enough" for a disease to become endemic) \cite{KermMcKen:27}. Since removal
  rate from the infective class is $\alpha$
  (in units of 1/time), the average period of infectivity is $1/\alpha $. Thus $\beta /\alpha$ is the fraction of the population that comes into contact with an infective individual during the period of infectiousness \cite{Edelstein:05}. The equilibrium
  $i=\gamma (n - s)/(\alpha + \gamma)$, which corresponds to $s=\alpha /\beta$, is called an endemic equilibrium.

   The basic reproduction number for a disease is the number of secondary infections produced by an infected individual in a population of susceptible individuals. It is a known fact that the basic reproduction number ${\cal R}_0$ is one of the most used thresholds in epidemic theory \cite{DriesWat:02}. It allows us to determine whether there is a rumor epidemic.

   Note that the matrix of the linearization of (\ref{SIRSmodel}) at an equilibrium $(s,i)$ is
   \begin{equation*}
    \left(
        \begin{array}{cc}
          -(\beta i + \gamma ) & -(\beta s + \gamma)  \\
          \beta i &  \beta s - \alpha \\
        \end{array}
      \right).
\end{equation*}
   By observing the sign structure at the disease-free equilibrium this matrix has negative trace and positive determinant
   if and only if $\beta n < \alpha$, or ${\cal R}_0 < 1$. At an endemic equilibrium, the sign structure shows
    that the matrix also has negative trace and positive determinant (see \cite{BauerCast:12} for details). In conclusion, the rumor-free equilibrium is asymptotically stable if and only if ${\cal R}_0 < 1$ and the endemic equilibrium, which exists if and only if ${\cal R}_0 > 1$, is always asymptotically stable, so that the rumor persists.

    Even if the endemic equilibrium is unstable, the instability commonly arises from a bifurcation and the infection (rumor) still
    persists but in an oscillatory manner \cite{BauerCast:13}. More specifically, as ${\cal R}_0$ increases through 1 there
    is an exchange of stability (bifurcation) between the rumor-free equilibrium and the endemic equilibrium. This transition
    is called a forward bifurcation. It has been noted (see references in \cite{BauerCast:13}) that in epidemic models with multiple interaction mechanisms it is possible to have a very different bifurcation behavior at ${\cal R}_0 = 1$.
    There may be multiple positive endemic equilibria for values of ${\cal R}_0 < 1$,  resulting in the so-called
     backward bifurcation at ${\cal R}_0 = 1$. For details see \cite[Ch. 2]{BauerCast:13}.

  Although at the beginning of a rumor outbreak there is a very small number of affected individuals, and the transmission of rumors is a stochastic event depending on the pattern of contacts between members of the population, we seek to define a deterministic model that simulates the rumor phenomenon in the near future. 
  In the following sections of this paper, we assume that we are in a rumor epidemic situation (i.e., a compartmental model), which follows a disease outbreak that probably was modeled (initially) by a stochastic process (i.e., a branching process \cite{BauerCast:12}).

  Finally, it is clear that contacts do not necessarily transmit the rumor. For each contact between infected
  and susceptible individuals, there is a probability that the infection will actually be transmitted. We assume that there is a mean probability $T$, called the {\it transmissibility} \cite{BauerCast:12}, of transmission of rumor. The transmissibility depends on the rate of contacts, the probability that a contact will transmit rumor, the duration time of the rumor, and the susceptibility. However, to simplify the model, in this article we will assume that all contacts transmit rumor (i.e., $T=1$).

    This paper is organized as follows.
    In Section~2, we make a brief presentation on some known mathematical models for the spread of rumors.
     The proposed deterministic PR model is presented in Section~3.
    Finally, in Section~4 we present some concluding remarks.

%% file: KnownModel.tex
\section{Some known PR models}
Rumors can be viewed as an ``infection of the mind" \cite{NekMorBia:07}. Therefore, the rumor propagation problem 
can be studied with many modeling techniques used in the study of epidemics. Some of these approaches include 
deterministic models, stochastic models and complex networks.

\subsection{Classical epidemic models of rumor spreading}
The DK model \cite{DaleyKendall:65} assumes that individuals in the network are categorized into three groups: {\it ignorants}
 ($s$), {\it spreaders} ($i$), and {\it stiflers} ($r$), such that $s$ represents people who are ignorant of the rumor, $i$ is people who actively spread the rumor, and $r$ is people who have heard the rumor, but no longer are interested in spreading it.
 (Sometimes here, we will use these terms interchangeably.)
 The rumor is diffused through the population by pairwise contacts between spreaders and others in the population. Any spreader involved in a pairwise meeting attempts to contaminate the other individual with the rumor, following the law of mass action. If this other individual is an ignorant, it becomes a spreader. If those involved in the meeting are a spreader and a stifler, either one or both of those, learn that the rumor is known and decide not to tell the rumor anymore, therefore turning into stiflers.

In the model proposed by Maki and Thompson \cite{MakiThomp:73} (MK model), the rumor is spread by direct contact of the spreaders with others in the population. In addition, when a spreader contacts another spreader, only the initiating spreader becomes a stifler. If we introduce the normalized variables $s/n$ and $i/n$ (denoted with the same symbol), and since $n=s+i+r$, the proposed model is as follows
 \begin{equation*} \label{MKmodel}
\left\{ \begin{array}{lll}
\frac{ds}{dt} & =  & -\beta si, \\[1.5mm]
\frac{di}{dt} & = & \beta si - \nu i^2 - \nu i(1-i-s)=(\beta + \nu )si - \nu i.
 \end{array}
 \right.
\end{equation*}
Compared with the simple SIR model (\ref{KMmodel}), we see that the only difference is that we have a factor $\beta + \nu$ in the second equation instead of $\beta $. It is clear also that the ignorant population is decreasing, nonnegative since $s,i\geq 0$ and
$ds/dt \leq 0$. Since ${\cal R}_0=(\beta + \nu)/ \nu$, ${\cal R}_0>1$ if and only if  $\beta /\nu >0$. I.e., there will be a rumor epidemic even for arbitrarily small rate parameters.

 Recently \cite{AllenLah:12}, for the epidemic model, summarize some of the deterministic and stochastic threshold theory, illustrate how to calculate the stochastic thresholds, and derive some new relationships between the deterministic and stochastic thresholds. We think that this approach can be applied to the PR model.

\subsection{Complex networks}
The so-called ``complex networks" have non-trivial topological features and can vary from technological networks to social networks and biological networks \cite{Erciyes:15}. The previous models assume a homogeneously mixing population and  do not take into account the topology of the underlying social interaction networks along which rumors spread. While such simple models may adequately describe the spreading process in small scale social networks, via the word-of-mouth, they become highly inadequate when applied to the spreading of rumors in large social interaction networks or by the Internet, which involves millions of nodes. The topology of such large social networks shows highly complex connectivity patterns \cite{NekMorBia:07}.

In 2004 Moreno {\it et al.} \cite{MorNekPac:04} derived the mean-field equations characterizing the dynamics of a rumor process that takes place on top of complex heterogeneous networks. These equations were solved numerically by means of a stochastic approach. Then, they studied the spreading process for random scale-free networks. Rumor spreading forms the basis for an important class of communication protocols, called gossip algorithms, which are used for large-scale information dissemination on the Internet, and in peer-to-peer file sharing applications \cite{NekMorBia:07}.

In \cite{NekMorBia:07}, Nekovee {\it et al.} made several contributions to the study of rumor dynamics on complex social networks. They introduce a new model that unifies the MK model with the SIR model of epidemics, and has both of these models as its limiting cases. They also describe a formulation of this model on networks in terms of Interacting Markov Chains (IMC), and use this framework to derive, from first-principles, mean-field equations for the dynamics of rumor spreading on complex networks with arbitrary degree correlations.

More recently, Cheng {\it et al.} \cite{Cheng:13} considered an online social site consisting of $n$ individuals which can be subdivided into three classes including  ignorants ($s$), spreaders ($i$), and stiflers ($r$). They introduced a stochastic epidemic model of the rumor diffusion. 
 The model assumes three status transitions in the model: from ignorants to spreaders, from spreaders to stiflers with contacts, and the spreaders to stiflers spontaneously. Unlike previous rumor diffusion models, this model treats the infectious probability as a variable, which can be seen as a function of strength of ties.

\subsection{Deterministic vs. stochastic models}
In this section the advantages and disadvantages of each type of model are pointed out.

It is interesting to note that most models describing epidemic spreading are deterministic because they require less data, are relatively easy to set up, and because the computer software to study them is widely available and user-friendly. ``In fact, models often identify behaviors that are unclear in experimental data-often because data are not reproducible and the number of data points is limited and subject to errors in measurement" \cite{BauerCast:12}. The dynamical behavior of deterministic models, also known as compartmental models, attempt to describe and explain what happens on the average at the population scale. They fit well large populations and are now well understood, so that deterministic models are generally the chosen strategy to apply when a new problem is posed.

On the other hand, stochastic models are based on probabilities rather than definite rates. There are much more complex models, such as those considered above in the study of complex networks, which generally involve stochastic procedures and provide much more insight into an individual-level modeling; however, they can be laborious to set up and need many simulations to yield useful predictions.

Here we propose a (non-stochastic) deterministic mathematical model for the spread of rumors, and we leave for future work
 the adaptation of our approach to the derivation of relationships between deterministic and stochastic thresholds \cite{AllenLah:12}.

%% file: PR_Model.tex
\section{Deterministic PR models} \label{determPRmodel}
Thompson {\it et al.} \cite{ThoCasDauCin:03} started from the same framework as DK model and applied a deterministic approach; however,
they considered elements such as the heterogeneity in the susceptible and spreader classes, as well as transmission, on the spread of a rumor $-$aspects not considered in the DK model. So they defined the passive and active people to be those who do not have many contacts and those who have many contacts, respectively. In this Section, a different approach, but also based on a deterministic mathematical model, is introduced.

\subsection{PR model with an exponentially distributed period of temporary immunity} \label{PRmodel1}
From the SIRS model (\ref{SIRSmodel}), we will investigate the effect of introducing a second rumor in the same population of individuals, which will try to counteract the effect of the original rumor. A situation that occurs frequently in communities, when a rumor is counteracted by a contrary information or news, which behaves in the same way as a rumor. The proposed model is inspired by the model of vaccination \cite{BauerCast:12}, which we add to a SIRS-type model (see (\ref{SIRSmodel})). In a way that we hope that vaccination (second rumor) reduces susceptibility to disease (i.e., the first rumor).

From the study carried out by Brauer and Castillo-Chavez in \cite[Ch. 2]{BauerCast:13} (see also \cite{Brauer:04}, \cite{Brauer:11})
to the case of a SIS model with vaccination,
 we propose here an adaptation to the problem of spreading rumors.
More specifically, we propose an SIRS-type model defined by system (\ref{SIRSmodel}), which assumes an exponentially distributed period of temporary immunity, after which the removed individuals (stiflers) can return to the susceptible class (ignorants). Furthermore, as in \cite[Ch. 2]{BauerCast:13}, we will consider the inclusion of a new class of individuals, the vaccinated ones ($v$). For simplicity, just as in epidemic models, we do not include births, natural deaths, and disease deaths \cite{BauerCast:13}, so that the total population size may be taken as constant. We will add the assumption that a fraction $\phi$ of the susceptible class knows the second rumor
 (i.e., it is vaccinated).
Also, 
 in this model we assume that individuals who know the second rumor have a susceptibility to first rumor reduced by a factor
 of $\sigma$ ($0\leq \sigma \leq 1$),
 where $\sigma =0$ corresponds to the situation where the second rumor counteracted the effect of the first rumor $100\%$, and
 $\sigma =1$ corresponds to the situation where the second rumor had no effect on the first one.
 Thus, the PR model is expressed as
\begin{equation} \label{SIRSVmodel}
\left\{ \begin{array}{lll}
\frac{ds(t)}{dt} & =  & -\beta s(t)i(t) - \phi s(t) + \gamma (n - i(t) - s(t)), \\[1.5mm]
\frac{di(t)}{dt} & = & \beta s(t)i(t) + \sigma \beta v(t)i(t) - \alpha i(t), \\[1.5mm]
\frac{dv(t)}{dt} & = & \phi s(t) - \sigma \beta v(t)i(t),
 \end{array}
 \right.
\end{equation}
subject to the initial conditions prescribe $s(0)$, $i(0)$, $v(0)$, with $s(0)+i(0)+v(0)=n$, and where $\gamma $, as in (\ref{SIRSmodel}), represents the proportional rate of loss of immunity. 
 The definition of the parameters involved can be seen in Table~\ref{paramdef}.
\begin{table}[htp!]
\caption{Parameters definition for the PR model.}  \label{paramdef}
\begin{center}
\begin{tabular}{c|c|c}
\hline
parameter & description & units  \tabularnewline
\hline
$n$  & population size (constant) & $(individuals)$ \tabularnewline
$\beta $  & infection rate & $(individuals \times time)^{-1}$  \tabularnewline
$\alpha $  & removal rate  &  $(time)^{-1}$  \tabularnewline
$\phi $  & fraction of the susceptible (ignorant) class    & $(time)^{-1}$  \tabularnewline
          &  to the first rumor which knows the    &      \tabularnewline
          &  second rumor per unit time                       &      \tabularnewline
$\gamma $  & proportionality rate of loss of  & $(time)^{-1}$  \tabularnewline
           &   immunity to the first rumor       &              \tabularnewline
$\sigma $  & susceptibility-reduction factor to first  &  $(individuals)^{-1}$ \tabularnewline
            &    rumor caused by the second rumor       &      \tabularnewline
\hline
\end{tabular}
\end{center}
\end{table}
 Since $n=s+i+v$ is constant, $s=n-i-v$, and the system (\ref{SIRSVmodel}) is equivalent to the system
 \begin{equation} \label{SIRSVequiv}
\left\{ \begin{array}{lll}
\frac{di(t)}{dt} & = & \beta (n - i(t) - (1-\sigma )v(t))i(t) - \alpha i(t), \\[1.5mm]
\frac{dv(t)}{dt} & = & \phi (n - i(t)) - \sigma \beta v(t)i(t) - \phi v(t).
 \end{array}
 \right.
\end{equation}
This is the basic model which we will study closely. Note that if all ignorants heard the second rumor immediately
($\phi \rightarrow \infty $),  there will be no ignorant individuals who have not been vaccinated, $n=i+v$, and
the model (\ref{SIRSVequiv}) is equivalent to
\begin{equation} \label{logistic}
\frac{di(t)}{dt}  =  \sigma \beta i(t)(n - i(t)) - \alpha i(t),
\end{equation}
which is an SIS model with basic reproductive number ${\cal R}_0^* = \sigma \beta n/ \alpha = \sigma {\cal R}_0 \leq {\cal R}_0$,
 where ${\cal R}_0$ is the basic reproductive number of the model (\ref{SIRSmodel}). But (\ref{logistic}) is also a logistic equation, which is easy to solve analytically or qualitatively. It coincides with the simple model proposed in \cite{Pearce}
 (at the beginning of Section~\ref{Introd}, paragraph two, it was briefly discussed). As it is usual in epidemiological models,
 the rumor-free equilibrium corresponds to $i = 0$ (so necessarily $v = n$), and is stable or unstable depending on the basic reproductive number \cite{BauerCast:12}.

 \subsubsection{Stability analysis}
 Following the strategy proposed in \cite[Ch. 2]{BauerCast:13} to the case of a SIS model with vaccination, we consider
 the parameters $\alpha $, $\beta $, and $\sigma $ as fixed and analyze the effect of varying $\phi$ (since in
practice this parameter is the one most easily controlled \cite{BauerCast:13}). We denote by ${\cal R}(\phi )$ the basic reproductive number
of the model defined by (\ref{SIRSVequiv}). Note that equilibria of this model are solutions of following system
\begin{eqnarray*}
\beta (n - i(t) - (1 - \sigma )v(t))i(t) & = & \alpha i(t) , \\
\phi (n - i(t)) & = & \sigma \beta v(t)i(t) + \phi v(t).
\end{eqnarray*}
Observe that if $i= 0$, then the first equation is satisfied and from the second we obtain $v=n$,
which corresponds to the rumor-free equilibrium. Now, the matrix of the linearization of (\ref{SIRSVequiv}) at an equilibrium
$(i ,v )$ and whose determinant we want to estimate is given by
 \begin{equation*}
\left( \begin{array}{cc}
-2 \beta i - (1 - \sigma )\beta v - \alpha + \beta n    &    -(1 - \sigma )\beta i \\
 -(\phi + \sigma \beta v)    &     -(\phi + \sigma \beta i)
 \end{array}
 \right).
\end{equation*}
And at the rumor-free equilibrium the matrix is
 \begin{equation*}
\left( \begin{array}{cc}
 -(1 - \sigma )\beta n - \alpha + \beta n    &    0 \\
 -(\phi + \sigma \beta n)    &     -\phi
 \end{array}
 \right),
\end{equation*}
which has negative eigenvalues (implying the asymptotic stability of the rumor-free equilibrium) if and only if
$-(1 - \sigma )\beta n + \beta n < \alpha $, which is equivalent 
 to $\sigma \beta n < \alpha $ 
 or to ${\cal R}(\phi ) = \sigma \beta n/ \alpha = \sigma {\cal R}_0 < 1$.

 Also note that
  \begin{equation*}
{\cal R}(\phi ) = \left\{ \begin{array}{ll}
\; {\cal R}_0,           &    \rm{ if } \; \phi =0 \; (\rm{there \; is \; no \; a \; second \; rumor}) \\
\sigma {\cal R}_0,    &   \rm{ if } \; \phi >0 \; (\rm{so}\; {\cal R}(\phi ) < {\cal R}_0 )
 \end{array}
 \right.
\end{equation*}
So ${\cal R}(\phi ) \leq {\cal R}_0 $. Following a similar analysis as in \cite{BauerCast:13}, we assume first that $0< \sigma < 1$
to get endemic equilibrium points from
\begin{eqnarray} \label{sist1} \label{sist1}
\beta (n - i - (1 - \sigma )v) & = & \alpha , \\  \label{sist2}
\phi (n - i) & = & \sigma \beta vi + \phi v.
\end{eqnarray}
We will do this by trying to find the unknown $v$ from the first equation of (\ref{sist1}) and substitute into the second equation
to get an equation of the form
\begin{equation}   \label{ecncuad}
ai^2 + bi + c = 0, \; {\rm where} \; a= \sigma \beta ,\; b=\sigma (\phi + \alpha - \beta n) \;\; {\rm and} \;\;
c=(\alpha \phi / \beta ) - \sigma \phi n .
\end{equation}

Note that if ${\cal R}(\phi ) < 1$, ${\cal R}(\phi ) = 1$, or ${\cal R}(\phi ) > 1$, then $c>0$, $c=0$, or $c<0$, respectively.
 Also, it is clear that if ${\cal R}(\phi ) > 1$, then there is a unique positive root of (\ref{ecncuad})
 and thus there is a unique endemic equilibrium. Now, if ${\cal R}(\phi ) = 1$ ($c=0$), there is a unique
 nonzero solution of equation (\ref{ecncuad}) $i=-b/a$, which is positive if and only if $b=\sigma (\phi - \alpha - \beta n) <0$.
  In this case there is a positive endemic equilibrium. Since equilibria depend continuously on $\phi $
  there must then be an interval to the left of ${\cal R}(\phi ) = 1$ in which there are two possible equilibria; these are
  $i=(-b \pm \sqrt{b^2 - 4ac})/2a $. This shows that the system (\ref{SIRSVequiv}) has a backward bifurcation
  (see Section~\ref{Introd}) at ${\cal R}(\phi ) = 1$ if and only if $b<0$ and $\beta $ is such that $c=0$.

  So, if $c>0$ and $b \geq 0$, there are no positive solutions of (\ref{ecncuad}), therefore there are no
 endemic equilibria. Equation (\ref{ecncuad}) has two positive solutions, corresponding
to two endemic equilibria, if and only if $c>0$ (${\cal R}(\phi ) < 1$) and $b<0$, $b^2 > 4ac$, or $b < -2 \sqrt{ac}$.
If $b = -2 \sqrt{ac}$, there is one positive solution $i=-b/2a$ of (\ref{ecncuad}).

   It can also be proved that such backward bifurcation is not possible for an SI model
 (case $\alpha =0$) \cite{BauerCast:13}.

 If $\sigma = 0$, from the equilibrium conditions (\ref{sist1})-(\ref{sist2})
 we cannot infer anything conclusive. However,
  from the more general epidemiological approach of \cite[Ch. 2]{BauerCast:13}, it follows that there is a unique endemic
equilibrium if ${\cal R}(\phi ) > 1$ and there cannot be an endemic equilibrium if ${\cal R}(\phi ) < 1$.
Therefore, in this case, it is not possible to have a backward bifurcation at
${\cal R}(\phi ) = 1$.

Alternatively, if ${\cal R}(\phi ) = 1$, $c = 0$, and $\sigma \phi \beta n = \alpha \phi $.
 Also, condition $b <0$ is equivalent to $\sigma (\phi + \alpha) < \sigma \beta n$. From whence it follows that
 $(\sigma \phi )^2 < \alpha \phi \sigma (1 - \sigma)$. Thus, a backward bifurcation occurs at ${\cal R}(\phi ) = 1$ if
 and only if this expression is satisfied. Therefore, a backward bifurcation is not possible if $\sigma = 0$ or $\sigma = 1$,
 or if $\alpha = 0$ (i.e., for an SI model).

\begin{figure}[htp!]
   \begin{center}
   \includegraphics[width=9.0cm]{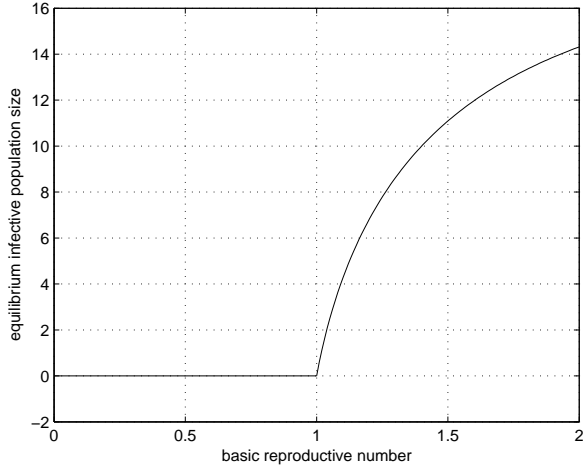}
   \caption {Forward bifurcation. As ${\cal R}(\phi )$ is a constant multiple of $\beta $, we can think of $\beta $
   as the independent variable in the bifurcation curve with the other parameters as constant \cite{BauerCast:13}.}
   \label{figure_fb}
   \end{center}
\end{figure}

\begin{figure}[htp!]
   \begin{center}
   \includegraphics[width=9.0cm]{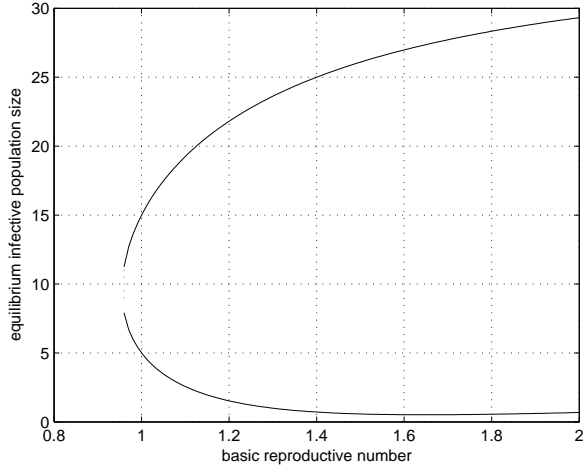}
   \caption {Backward bifurcation.}
   \label{figure_bb}
   \end{center}
\end{figure}

As ${\cal R}(\phi )=(\sigma n/\alpha )\beta $, for a glimpse of how the bifurcation curve is in the $({\cal R}(\phi ),i)$-plane,
 we can consider $ \beta $ as the independent variable and all other parameters as constants.
 Differentiating implicitly with respect to $\beta $ the equation (\ref{ecncuad}) we obtain
 \[
 (2ai + b)\frac{di}{d\beta } = \sigma i(n-i) + \frac{\alpha \phi }{ \beta ^2}.
\]
From the equilibrium condition (\ref{sist2}) we infer that $i\leq n$, which means that the
bifurcation curve has positive slope at equilibrium values with $2ai + b >0$ and negative
slope at equilibrium values with $2ai + b < 0$. If there is not a backward bifurcation at
${\cal R}(\phi ) = 1$, then the unique endemic equilibrium for ${\cal R}(\phi ) > 1$ satisfies $2ai + b = \sqrt{b^2 - 4ac} >0$,
 and the bifurcation curve has positive slope at all points where $i > 0$. The bifurcation curve is illustrated in Figure~\ref{figure_fb}, using parameter values shown in Table~\ref{paramvalest} ($\beta $ being non-constant).
 If there is a backward bifurcation at ${\cal R}(\phi ) = 1$, then there is an interval on which there are two endemic equilibria given by $2ai+b = \pm \sqrt{b^2-4ac}$, and the bifurcation curve has negative slope at the smaller of these and positive slope at the
larger of these (see Figure~\ref{figure_bb}).

 Note that from the matrix of the linearization of (\ref{SIRSVequiv}) at an equilibrium $(i ,v)$ and using the equilibrium conditions
 (\ref{sist1})-(\ref{sist2}), we can get the matrix at an endemic equilibrium $(i ,v)$; this is
  \begin{equation*}
\left( \begin{array}{cc}
 - \beta i    &    -(1- \sigma )\beta i \\
 -(\phi + \sigma \beta v)    &   -(\phi + \sigma \beta i)
 \end{array}
 \right).
\end{equation*}
 This matrix has negative trace and determinant equal to $\beta i (2ai + b)$. So, if $2ai + b > 0$ (i.e., if the bifurcation curve has positive slope), then the determinant is positive and the equilibrium is asymptotically stable. If $2ai + b < 0$,
 the determinant is negative and the equilibrium is unstable (i.e., it is a saddle point). So that an endemic equilibrium
 of PR model (\ref{SIRSVequiv}) is (locally) asymptotically stable if and only if it corresponds to a point on the bifurcation curve at which the curve is increasing. For details of a similar analysis, see \cite[Ch. 2]{BauerCast:13}.

 \subsubsection{Parameter estimation and numerical simulations}
 Clearly, the parameters may vary depending on the type of person, the situation in which the rumor is transmitted,
 and the quality of the rumor itself. We have, basically, two types of rumor which determine the population classes involved:
 susceptibles or ignorants, infectious individuals or spreaders, and 
 those infected with a second rumor or vaccinated individuals
 which, without being spreaders, will seek to counteract the effects of the first rumor. Let us also remember that those individuals who lost interest in the original rumor (i.e., the removed individuals or stiflers) have temporary immunity (SIRS model).
  So that these, along with those who know the second rumor (vaccinated), but that are susceptible to becoming contaminated with the
  first rumor (reduced by a factor of $\sigma $), are also part of the susceptible population (ignorants).

In our model, which involves temporary immunity and two kinds of rumors, we propose to model the interaction between
 individuals on the Internet and social networks. This choice is justified by some statistical information available in
 Internet. This is:
\begin{itemize}
\item Almost half the world's population have internet access and nearly one-third of the world's population now uses social media.
\item The average user checks his social network at least twice a day. Social networks are the number 1 activity on Internet.
\item 80\% of social media users interact in some way with other users or profiles.
\item On average, active users have 200 to 400 friends and receive more messages than they send.
\end{itemize}

Following the methodology of estimating parameter values proposed in \cite{ThoCasDauCin:03}, we considered working with
these means of transmission because we can get updated information on their use and there is a lot of data online.
 We consulted the information from websites \cite{TWBank1} and \cite{TWBank2} to approximate the number
 of users of the means of transmission; and thus, to estimate how many interactions occur on average, which allowed us
 to approximate the values of the parameters that we need. See Table~\ref{paramvalest}.
 \begin{table}[htp!]
\caption{Parameter values estimated for the PR model.}  \label{paramvalest}
\begin{center}
\begin{tabular}{c|c}
\hline
Parameters &  Values  \tabularnewline
\hline
$\beta $  &   $0.9$ \tabularnewline
$\alpha $    &  $0.3$  \tabularnewline
$\phi $      & $0.7$  \tabularnewline
$\gamma $     & $0.2$  \tabularnewline
$\sigma $    &  $0.2$ \tabularnewline
\hline
\end{tabular}
\end{center}
\end{table}

\begin{figure}[htp!]
    \begin{center}
   \includegraphics[width=17.5cm]{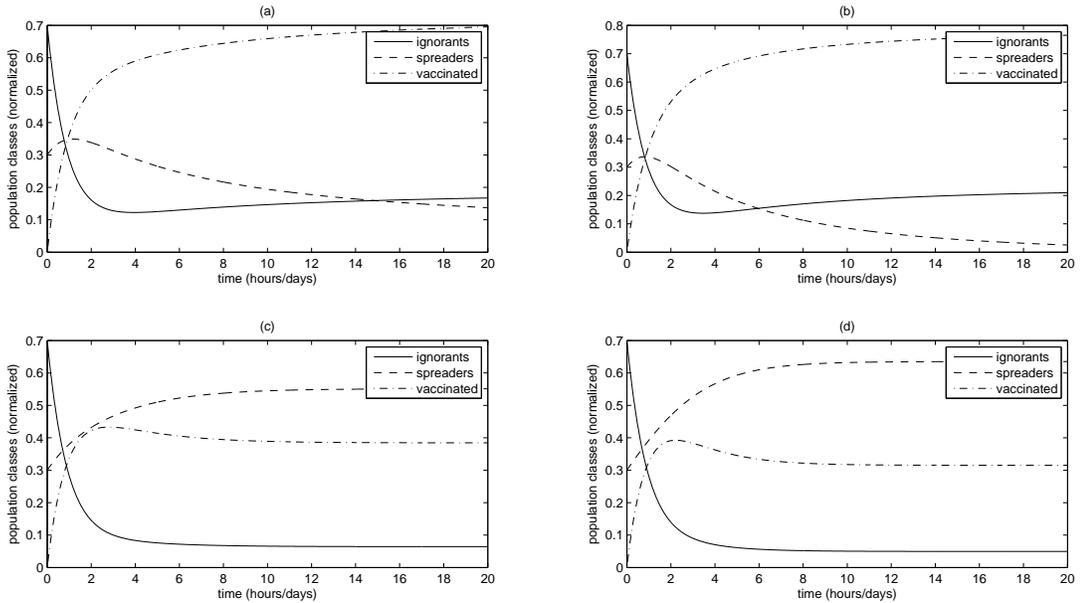}
   \caption {Behavior simulations for the PR model (\ref{SIRSVmodel}) for different $\sigma $ parameter values: (a) Using
   Table~\ref{paramvalest}. (b) Case $\sigma =0$, which corresponds to the situation where the second rumor counteracted the effect of the first rumor $100\%$. (c) Using $\sigma = 0.7$. (d) Case $\sigma = 0.9$, which corresponds to the situation where the second rumor ``almost" had no effect on the first one.}
   \label{figure_siv}
     \end{center}
\end{figure}

Figure~\ref{figure_siv} shows, in the same coordinate plane, the curves corresponding to ignorants or susceptibles ($s(t)$),
infectives or spreaders of the first rumor ($i(t)$), and spreaders of the second rumor or vaccinated individuals ($v( t)$),
for different $\sigma $ parameter values. Observe the typical behavior of infectives, according to which the number of infected individuals increases sharply and then slowly decreases until it disappears. We also note regarding the behavior of susceptible individuals that during the time of spreading the rumor, its population decreases, but it does not disappear, instead it maintains a
slight but steady increase. Finally, for values of $\sigma $ equal to $0.2$ and $0.0$ (Figures~\ref{figure_siv}(a)
 and~\ref{figure_siv}(b) respectively) the second rumor, as time passes, increases rapidly at first and then more slowly but steadily, thus counteracting the effect of the first rumor. If $\sigma $ takes a value close to $1$ (for example, 0.7 or 0.9, see Figures~\ref{figure_siv}(c)
 and~\ref{figure_siv}(d)), this situation is reversed.
\subsection{An alternative simple model for the propagation of two conflicting rumors} \label{PRmodel2}
Now, we consider heterogeneity in behavior, specifically contact rates. We will describe the case of
two conflicting rumors spreading over a population of individuals of size $n$.
The proposed model is based on the vaccination model described in \cite[Ch. 9]{BauerCast:12}, \cite[Ch. 3]{BauerCast:13},
 under the assumption that vaccination (the second rumor) reduces susceptibility to the first rumor.
Our PR model requires considering two sub-populations, the first one knows the first rumor and the other knows the second
rumor. For clarity, as in the case of the vaccination model, we assume that a fraction of the population knows
the second rumor (i.e., individuals who have been vaccinated) prior to the 
spreading of the first rumor, which produces the first rumor-infected individuals, who are not necessarily infective.
We will describe a PR model with compartmental SIR structure in both sub-populations.
For simplicity, as before, we do not include births, natural deaths, and disease-related deaths. The two sub-populations have
constant total sizes $n_1$ and $n_2$, and $n_1+n_2=n$. We assume mass-action contact with contact rates
$\beta_1$, $\beta_2$ and recovery rates $\alpha_1$, $\alpha_2$, respectively. We assume also that those individuals
who know both rumors have reduced their ability to spread the first rumor by a
factor of  $\eta $. As in Section~\ref{PRmodel1}, we also assume that individuals who
know the second rumor have susceptibility to first rumor reduced by a factor $\sigma$  ($0\leq \sigma \leq 1$).
 The proposed PR model is as follows:
   \begin{equation} \label{twoSIRmodel}
\left\{ \begin{array}{lll}
\frac{ds_1(t)}{dt} & =  & -\beta_1 s_1(t)i_1(t) -\beta_1 s_1(t)\eta i_2(t) = -\beta_1 s_1(t)[i_1(t) + \eta i_2(t)], \\[1.5mm]
\frac{di_1(t)}{dt} & = & \beta_1 s_1(t)i_1(t) + \beta_1 s_1(t)\eta i_2(t) - \alpha_1 i_1(t) =
\beta_1 s_1(t)[i_1(t) + \eta i_2(t)]- \alpha_1 i_1(t), \\[1.5mm]
\frac{ds_2(t)}{dt} & = & - \sigma \beta_2 s_2(t)i_1(t) - \sigma \beta_2s_2(t)\eta i_2(t) =
- \sigma \beta_2 s_2(t)[i_1(t)+\eta i_2(t)],  \\[1.5mm]
\frac{di_2(t)}{dt} & = & \sigma \beta_2 s_2(t)i_1(t) - \sigma \beta_2s_2(t)\eta i_2(t) - \alpha_2 i_2(t) =
\sigma \beta_2 s_2(t)[i_1(t)+\eta i_2(t)] - \alpha_2 i_2(t),
 \end{array}
 \right.
\end{equation}
 where $s_1(t)$, $i_1(t)$, $s_2(t)$, and $i_2(t)$ denote the number of first rumor susceptibles, the number of
 first rumor infectives, the number of second rumor susceptibles, and the number of second rumor infectives
 at time $t$, respectively. Likewise, $s_1(0)$, $i_1(0)$, $s_2(0)$, and $i_2(0)$ are the number of susceptibles and infectives at the initial time $t= 0$, with $s_1(0)+i_1(0) = n_1$ and $s_2(0)+i_2(0) = n_2$.


In this case, the infection by the first rumor is beginning in a population which is not fully ignorant.
 So we should speak of the control reproduction
 number ${\cal R}_c$ rather than the basic reproduction number \cite{BauerCast:13}. Through the introduction of the next generation matrix with large domain concept for system~(\ref{twoSIRmodel}) at the disease-free equilibrium it is relatively easy to see that $${\cal R}_c = \frac{\beta_1 n_1}{\alpha_1 } +  \frac{\eta \sigma \beta_2 n_2}{\alpha_2}$$ (see \cite{BauerCast:13} for details).
 We note that the basic reproduction number of the disease is the sum of the reproduction numbers for each group \cite{DriesWat:02}.
 So it should follow that the disease-free equilibrium (from the first rumor) is asymptotically stable if the reproduction number is less than 1 and unstable if it is greater than 1.
\subsection{A numerical example} \label{numexa}
\subsubsection*{Selecting the dataset}
We selected two rumors, from which the primary rumor to be studied is ``earth is flat" in contrast to ``earth is round". This topic was chosen because it is widely known that the Earth's shape is round, however, there are many internet users that have been spreading out the theory that the Earth is flat. This started first as a joke but shortly after, many people began supporting conspiracy theories about the flatness of the Earth.

The measured data was taken from the Google Trends open source platform, which allows any user to look up the statistics of a search term using the Google engine throughout time. The dataset we considered for this work has the daily search percent from the terms ``earth is flat", ``earth is round" and ``earth is" from February 18th, 2017 to March 17th 2017. The latter was taken into account so that we could study the rumors involving the interest on the Earth's features. Finally, ``earth is flat" and ``earth is round" search percent curves are relative to the ``earth is" search.

\subsubsection*{Numerical simulations}
Numerical simulations were performed using both the PR model with an exponentially distributed period of temporary immunity (original PR model) and the alternative simple model for the propagation of two conflicting rumors (alternative PR model).

Since the parameters of the proposed models are bounded, MATLAB's \verb+fminsearch+ function could not be used, because it was designed to find local minima of unconstrained multivariable functions \cite{mworks1}. Instead, we used the MATLAB's \verb+fmincon+ function, which is primarily used to solve nonlinear bounded optimization problems \cite{mworks2}. This MATLAB function has many implemented algorithms, such as: ``Interior-Point Optimization", ``SQP and SQP-Legacy Optimization", ``Active-Set Optimization", and ``Trust-Region-Reflective Optimization" \cite{mworks2}.
 By default \verb+fmincon+ will try to run the latter algorithm, but it requires the gradient of the objective function (which is very hard to calculate). Instead, it runs the ``active-set" algorithm described in \cite{mworks3}.

\begin{figure}[htp!]
    \begin{center}
   \includegraphics[width=10.5cm]{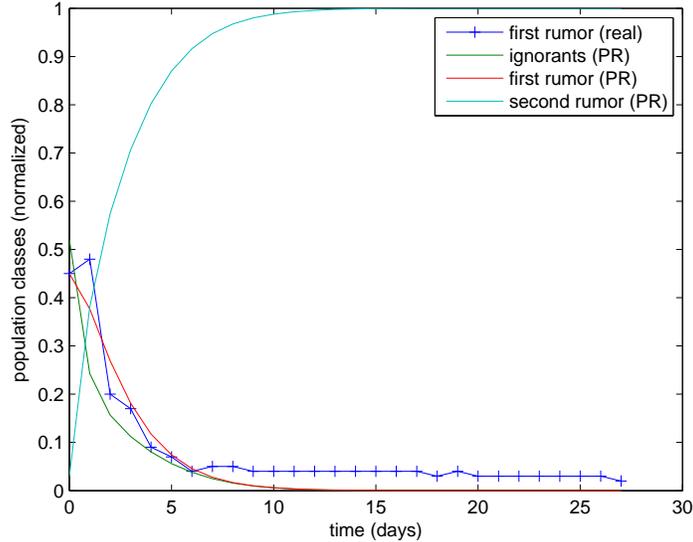}
   \caption {Behavior simulations for the original PR model (Section~\ref{PRmodel1}).}
   \label{Expormo}
     \end{center}
\end{figure}

 The initial point for the spreaders of the first rumor was $i(0)=0.45$ (according to the ``earth is flat" rumor) and for the second rumor, $v(0)=0.03$ (according to the number of users that searched ``earth is round" at the beginning of the spread of ``earth is flat"). Figure~\ref{Expormo} shows the original PR model's behavior.
 For the initial parameters $\alpha (0) = 0.3$, $\beta (0) = 0.9$, $\phi (0) = 0.5$, and $\sigma (0)=0.0$, the estimated parameters
 obtained for the simplified model~(\ref{SIRSVequiv}) are:
 \[  \alpha^* = 0.58,\quad \beta^* = 1.0, \quad \phi^* = 1.0, \quad \sigma^*=0.0.   \]
 These parameters suggest that the rumor is very easy to spread, however, the recovery rate is very fast because everybody knows that the ``earth is round". Also, this vaccine is strong enough to counteract the ``earth is flat" rumor in 100\%.

 However, as it is seen in Figure~\ref{Expormo}, the disease has not completely disappeared, contrary to the real data.
 A better fitting for this situation is provided by the following alternative PR model.

 \subsubsection*{Alternative PR model}
For the initial parameters $\beta_1(0) = 0.9$, $\alpha_1(0) = 0.9$, $\beta_2(0) = 0.5$, $\alpha_2(0) = 0.5$,
$\eta(0)=0.5$, and $\sigma(0)=0.5$, the estimated parameters obtained for the alternative model (\ref{twoSIRmodel}) are:
 \[  \beta_1^* = 0.1954, \quad \alpha_1^* = 0.4455,\quad \beta_2^* = 0.2825, \quad \alpha_2^* = 0.0, \quad \eta^* = 0.8428, \quad \sigma^*=0.2825.   \]
\begin{figure}[htp!]
    \begin{center}
   \includegraphics[width=17.5cm]{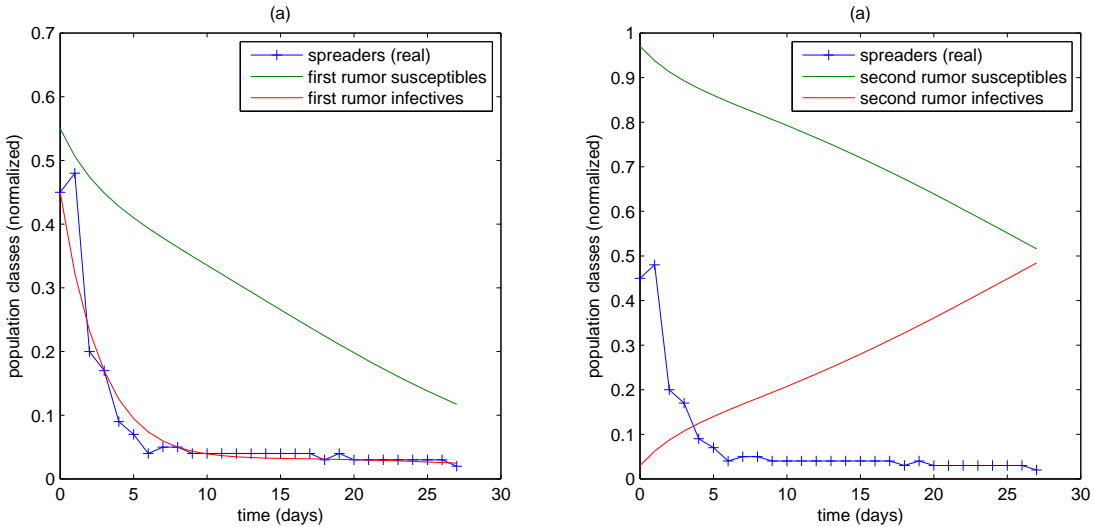}
   \caption {Behavior simulations for the alternative PR model (Section~\ref{PRmodel2}).}
   \label{Expalmo}
     \end{center}
\end{figure}
As it is seen in Figure~\ref{Expalmo}(a), the model fits accurately the data. The second rumor susceptible and infective curves are represented separately in Figure~\ref{Expalmo}(b) to make them easier to visualize. Finally, the SSE of the original PR model was $0.0398$, whereas the SSE of the alternative PR model was $0.0293$, so the latter in fact offers a more accurate fitting to the real data.

%% file: Conclus.tex
\section{Final remarks}
In this paper different possible behaviors in the dynamics of the rumor spreading has been studied.

In Section~\ref{PRmodel1} we proposed a deterministic PR model and detected relevant conditions, derived from local stability analysis
 of the rumor-free equilibrium and the rumor-endemic equilibrium. An important fact is that the endemic equilibrium is not
 asymptotically stable for all values of the parameters involved.
  However, 
  this model may have solutions that behave periodically \cite{BauerCast:12}. This means that the spread of a rumor still persists but in an oscillatory manner with possible variations in the number of ``contagions" as well as long periods of recurrence, and this makes its study and control difficult.
  A relevant problem to address is to identify situations of unstable endemic equilibrium for the spread of rumors.

   A backward bifurcation at $R_0=1$ makes the disease control more difficult. We corroborate the plausible fact that a
   possible measure to control the spread of a rumor could be to reduce the susceptibility of the population by spreading
   a second rumor (vaccination), which seeks to counteract the effect of the first one (the original infection), thus reducing
   the susceptibility of the population to the original rumor. Another measure could be to carry out a program of reeducation to ``convince" people not to express agreement with the rumor; alternatively, this strategy could also be seen as the introduction
   in the population a second rumor that tries to counteract the effect of the first one.

   In addition, in Section~\ref{PRmodel2} we proposed an alternative simple model, which provided a better fit to the data than did
   the original PR model in the numerical example considered in Section~\ref{numexa}.

   In the case of more than two rumors, we can assume that we are
concerned by only one of these rumors and the rest are
considered as contrary rumors. Hence, the problem can be
reduced to two conflicting rumors spreading over the same
population, one for the first rumor and the other englobing all
the adversary rumors. Thus, the results of this work can be extended to
the case of more than two rumors.

 As far as we know, the approach outlined here, using and adapting a compartment model to study the problem of spreading
 rumors, has not been stated before. This PR model confirms the parallelism between the study of the spread of a rumor and epidemiological study of the spread of an infectious disease. We think that it will be relatively
 easy to find direct applications for the results here considered. \\[3mm]



\noindent {\bf Acknowledgements:}\\
This research was partially supported by the Decanato de Investigaci\'on y Desarrollo (DID) at USB.